\documentstyle[aas2pp4]{article}
\lefthead{Steigman and Tkachev}
\righthead{$\Omega_{\rm B}$ and $\Omega_0$ From MACHOs}

\begin{document}

\input epsf

\def\alt{\mathrel{\mathpalette\fun <}}
\def\agt{\mathrel{\mathpalette\fun >}}
\def\fun#1#2{\lower3.6pt\vbox{\baselineskip0pt\lineskip.9pt
        \ialign{$\mathsurround=0pt#1\hfill##\hfil$\crcr#2\crcr\sim\crcr}}}

\title{$\Omega_{\rm B}$ and $\Omega_0$ From MACHOs and Local Group 
Dynamics}
\author{G. Steigman$^1$ and I. Tkachev$^{1,2,3}$}
\affil{
${}^1$Departments of Physics \& Astronomy, The Ohio State University\\ 
Columbus, OH 43210\\ 
${}^2$Institute for Nuclear Research of the Academy of Sciences of 
Russia\\ 
Moscow 117312, Russia\\
${}^3$TH Division, CERN, CH-1211 Geneva 23, Switzerland 
\footnote[4]{Current address}\\
}

\begin{abstract}
We obtain restrictions on the universal baryon fraction, $f_{\rm B} 
\equiv \Omega_{\rm B}/\Omega_0$, by assuming that the observed 
microlensing events towards the Large Magellanic Cloud are due to 
{\em baryonic} MACHOs in the halo of the Galaxy and by extracting 
a bound to the total mass of the Milky Way from the motion of tracer 
galaxies in the Local Group.  We find a lower bound $f_{\rm B} > 
0.29^{+0.18}_{-0.15}$.  Consistency with the predictions of primordial 
nucleosynthesis leads to the further constraint on the total mass 
density, $\Omega_0 \alt  0.2$.

\end{abstract}


%\twocolumn

\newpage

\section{Introduction}
It is a Herculean task to inventory the contents of the Universe 
(e.g., \cite{fhp}).  A more modest goal might be to pin down the 
baryonic fraction of the total mass, $f_{\rm B}$ (e.g., \cite{wnef}, 
\cite{xrc}).  If objects can be identified which are likely to provide
a ``fair" sample of $f_{\rm B}$, we may avoid the daunting prospect
of having to identify all the guises baryons may assume.  Large clusters
of galaxies offer a very promising site (\cite{wnef}; \cite{xrc}; 
\cite{xray1}; \cite{xray2}).  To test the estimates of the systematic
errors in $f_{\rm B}$ derived from X-ray cluster data, it would be 
of value to measure $f_{\rm B}$ in a completely different system, 
provided a case could be made that it will provide a ``fair" sample.  
Suppose, for example, we could estimate the baryonic mass associated
with the Galaxy.  If we could also measure the corresponding ``dynamical"
mass, we could obtain an independent estimate of $f_{\rm B}$ whose
systematic uncertainties (and dependence on the Hubble parameter) 
differ from those which accompany the X-ray cluster determinations.
In this paper we focus on the Local Group of galaxies (LG), using the 
MACHO mass estimates (\cite{MACHO}) for a lower bound on the baryonic 
mass and relying on LG dynamics to constrain the total mass estimate.

Microlensing experiments (\cite{MACHO}) suggest that roughly half the 
mass in the halo of our Galaxy, out to the distance of the Large 
Magellanic 
Cloud (LMC), may be in the form of Massive Compact Halo Objects (MACHOs).
One can imagine several exotic possibilities for the nature of the 
MACHOs. They could be very dense clusters of non-baryonic dark matter 
with special 
properties that allow them to clump inside their Einstein ring radii 
(\cite{kt94}), or they could be primordial black holes.  Neither of these 
possibilities is especially well motivated and each has its intrinsic 
difficulties, but neither can be excluded a priori.  Stellar remnants 
such as old white dwarfs\footnote{Neutron stars and black holes of 
stellar origin cannot constitute a significant halo fraction in view 
of the constraints arising from the observed metallicity and helium 
abundances (\cite{ros90}).} appear to offer a more natural candidate 
(\cite{MACHO}) which, however, is not without its problems too [e.g., 
white dwarfs require a rather narrow initial mass function in order 
to avoid overproducing low-mass stars or supernovae (\cite{AL96})].  
Dense and cold baryonic gas clouds have also been considered as a 
viable alternative for the observed gravitational microlenses 
(\cite{cbc}; 
\cite{cbc2}).  Finally, it must be kept in mind that the observed 
microlensing may be due to objects which are not in the halo of the 
Galaxy.  If the MACHOs are, indeed, stellar remnants (or cold baryonic 
gas clouds) in the halo of the Galaxy, then the mass of baryons within 
50 kpc of the Galactic center is $M_{\rm B} (50~{\rm kpc}) \geq M_{\rm 
MACHO} = 2.0^{+1.2}_{-0.7} \times 10^{11}M_\odot$ (\cite{MACHO}).

The purpose of the present paper is to extract information on the 
universal baryon fraction from this number assuming the MACHOs 
are revealing baryonic matter in the Galaxy halo, and from the 
dynamics of the Local Group of galaxies.  The constraint we obtain 
may be compared to the one derived from X-ray galaxy clusters (see, 
e.g., \cite{xrc}; \cite{xray1}; \cite{xray2}), but it relies on 
different observations in a completely different physical system 
on a vastly different scale and, interestingly, has a different 
dependence on the Hubble parameter ($H_0 \equiv 100h $~km~s$^{-1}
$~Mpc$^{-1}$).

The value of $M_{\rm B} (50~{\rm kpc})$ derived from microlensing 
experiments is approximately 50\% of the total mass of the Galaxy 
out to this distance.  The latter mass, presumably the sum of baryons 
and cold dark matter, is derived dynamically (see, e.g., \cite{koch}).  
However, on the basis of this we cannot conclude that the primordial 
baryon fraction is $f_{\rm B} \approx 0.5$.  Baryons are ``strongly'' 
interacting particles, while for the (non-baryonic) cold dark matter 
all interactions except gravitational can be neglected.  Consequently, 
the density profile of the baryonic matter does not necessarily follow 
the density profile of the cold dark matter, and baryonic matter may 
be more (or less) concentrated towards the 
center of the gravitational well.  However, we may be able to estimate 
the primordial baryon fraction if we take the ratio of baryons (as 
revealed by the MACHOs) to the total mass on some larger scale, which 
should be sufficiently large so that the matter inflow or outflow across 
the boundary of the region is negligible.

The total mass of matter residing in such a larger region can be found 
dynamically; however, we cannot measure the mass of baryons separately
on such larger scales.  Although the baryonic halo may be expected to 
extend outside of the 50 kpc scale (in the form, e.g., of MACHOs, diffuse 
gas, satellite Galaxies, etc.), by neglecting these extended baryons we 
can obtain a lower bound on $f_{\rm B}$.  Indeed, while in the past there 
might have been violent processes of baryon ejection from the Galaxy 
accompanying, e.g., supernova explosions, analogous ejecta of cold dark 
matter is not expected.  Therefore, by neglecting the unknown ejected 
component of baryons we will be on the ``safe side'' in our inequality 
for $f_{\rm B}$, which, we emphasize, does rely on our assumption that 
MACHOs are baryonic matter in the halo of the Galaxy.

For the larger reference scale we can choose the current turnaround 
radius for the  LG.  Initially, every shell of the Galaxy's building 
material expands with the Universe.  Gradually, this expansion slows 
down and eventually a gravitationally bound shell separates from the 
general expansion.  This shell stops expanding and then collapses 
(\cite{gg72}).  The radius of this first stopping point is the turnaround 
radius.  With the passage of time shells that are more and more distant 
and less and less bound turn around sequentially, i.e., the turnaround 
radius propagates outward with time (for details see, e.g.,  \cite{si}; 
\cite{si2}; \cite{stw}, 1997).  There is one shell that is turning 
around now, at present; the corresponding distance of this shell from 
the center of mass of the system is the current turnaround radius.  
Collisionless cold (non-baryonic) dark matter is restricted to remain 
within this radius, which is just what we want for the larger reference 
scale.  This picture of infall is valid independent of the assumption of 
spherical symmetry (the turnaround sphere will become a turnaround 
surface); 
for the model to be tractable analytically, we do assume spherical infall.

\section{Spherical Infall Model}

Let $R$ be the current turnaround radius, $M_{\rm B} (R)$ the mass 
of all baryons currently inside this radius, and $M_{\rm tot} (R)$ 
the total mass within $R$.  We are using the ratio $M_{\rm B} (R)/
M_{\rm tot}(R)  = f_{\rm B}$ to provide a measure of the universal 
baryon fraction, $f_{\rm B}= \rho_{\rm B}/\rho_{\rm tot}= \Omega_
{\rm B}/\Omega_{\rm 0}$.  For $R \geq 50$~kpc, we expect $M_{\rm B}(R) 
\geq M_{\rm B} (50~{\rm kpc})$.  Furthermore, there may be more baryons 
that were initially associated with the Galaxy than those that are 
currently within $50~{\rm kpc}$.  As a consequence, our estimate 
provides a {\it lower} bound on the universal fraction of baryons:
$f_{\rm B} > M_{\rm B} (50~{\rm kpc})/M_{\rm tot} (R)$.
 
Actually, in computing this inequality we do not have to use the 
current turnaround radius.  It is equally legitimate, and will 
provide a tighter constraint, to take any smaller radius that satisfies 
the following condition: matter which is falling freely and is 
currently at this radius has not yet had a chance to cross previously 
collapsed material.  In other words the shell is outside the first 
caustic of the spherical infall model.  The position of the first 
caustic is model dependent, but it has its largest value relative 
to the turnaround radius if the initial overdensity causing the 
infall can be considered as a point mass excess.  Let us denote 
the radius of the first caustic as $R_1$ and the total mass inside 
of it as $M_1$.  In the point mass excess case (and for $\Omega_0=1$), 
$R_1 = 0.37 R$ and $M_1 =0.7 M(R)$, (\cite{stw2}).

Before shell crossing the radius of any given shell obeys the equation 
of motion
\begin{equation}
\frac{1}{2}\dot{r}^2 - \frac{GM}{r}=-E \,\, ,
\label{eqm}
\end{equation}
where $M$ is the mass interior to the shell and $E$ is its binding 
energy per unit mass;
both are constants prior to shell crossing.  Solutions can be 
parameterized as
\begin{eqnarray}
r&=& A(1-\cos \theta) \, ,  \nonumber \\
t&=&B(\theta -\sin \theta) \, ,
\label{sol}
\end{eqnarray}
where $A \equiv GM/2E$, $B \equiv GM/(2E)^{3/2}$.  

Small galaxies in the 
Local Group that are close to the turnaround sphere can 
be considered as tracers of the motion of the corresponding shell.  If 
the distances to such galaxies and their radial velocities are known at 
time $t$, we can write, according to equations (\ref{sol}):
\begin{eqnarray}
\frac{vt}{r}&=& \frac{\sin \theta(\theta -\sin \theta)}
{(1-\cos \theta)^2} \, ,   \\
M&=&\frac{v^2 r}{G(1+\cos \theta)}\, .
\label{sol2}
\end{eqnarray}
Using equation (3) we can solve for $\theta$ which is then used in equation 
(4) to determine the total mass $M$ interior to the shell on which the 
tracer galaxy resides.  
% changes were made starting from here.
Note that the derivation of the relation $M = M(r,v,t)$ does not rely 
upon the initial mass distribution  (in other words we do not need to 
know the function $E = E(r_0)$).  In the spherical infall approximation 
$M$ gives a direct ``weighing'' of the LG at each radius where a known 
satellite galaxy resides. We shall use this procedure for a subset of 
galaxies that are close to the turnaround surface.

On the other hand, suppose that the initial mass distribution for 
unperturbed Hubble flow is known, e.g., $M(r_0) = 4\pi \rho r_0^3/3 
+ \delta M(r_0)$, where $\delta M(r_0)$ is an (initially small) excess 
mass over homogeneous cosmological background. A good example corresponds 
to the assumption of all excess mass concentrated at the origin, $r_0 =0$ 
(point mass excess). Using the integrals of the equations of motion written 
as $M(r,v,t) = M(r_0(r))$, we can find the velocity field at any given 
point, $v = v(r,t)$. The trajectory in the \{$r$, $v$\} phase space, 
$v=v(r)$, representing occupied cells at a given moment of time is called 
the infall trajectory.  For example, if the initial overdensity is scale 
independent, $\delta M /M = (M_0 /M)^\epsilon$, and for $\Omega =1$, the 
infall trajectory is given by
\begin{eqnarray}
r &=& R \,( \frac{1-\cos \theta}{2}) 
\left(\frac{\pi}{\theta -\sin \theta}\right)^{2/3+2/9\epsilon} \, , \\
v &=& \frac{r}{t} \, \frac{\sin \theta(\theta -\sin \theta)}
{(1-\cos \theta)^2} \, ,  
\label{infall_trj}
\end{eqnarray}
The point mass excess case corresponds to $\epsilon = 1$.

{\plotone{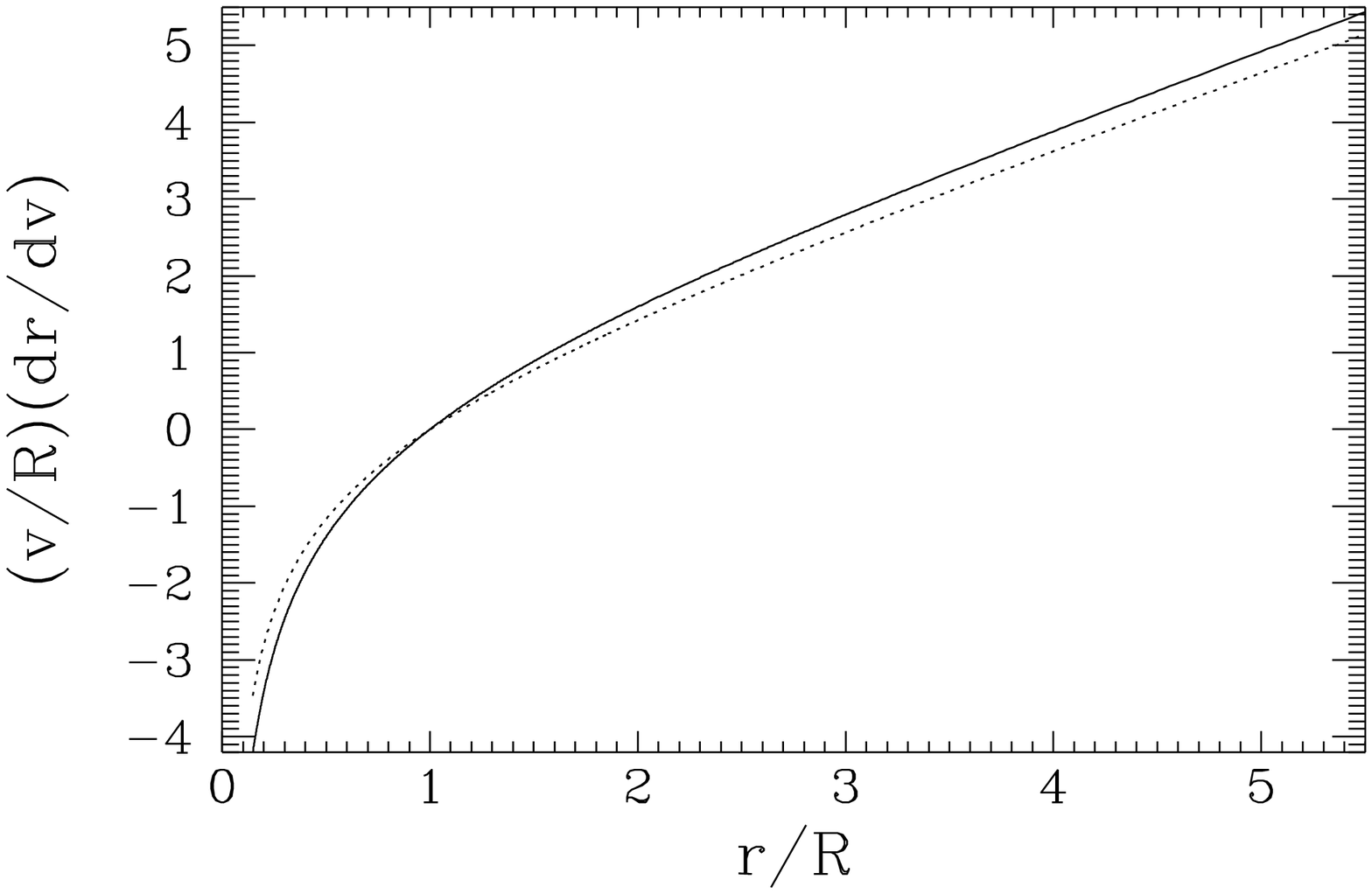}
\figcaption{
\baselineskip0.08cm
Infall trajectories for the cases $\Omega =1$ (solid line) and
$\Omega \rightarrow 0$ (dotted line).
\label{rv_omega}}
}

In the case $\Omega < 1$ the analytical expressions which give an infall
trajectory are more complicated and we do not present them here.  It is 
important, however, that at large $r$ the overdensity can be neglected 
and all infall trajectories tend to the simple Hubble law, $v = H r$. It 
is convenient to choose as physical parameters which characterize an infall 
trajectory the turnaround radius, $R$, and the Hubble constant, h, which 
is the slope of $v(r)$ at $r \gg R$.  When $R$ and $h$ are are fixed, 
the infall trajectories that correspond to different cosmological models, 
or different excess mass profiles, differ insignificantly outside of the 
first caustic. For comparison we plot infall trajectories for the case 
$\Omega =1$, $\epsilon = 1$ and $\Omega \rightarrow 0$, $\epsilon = 1$ in 
Fig.~1.  The difference between the two cases is significantly smaller than 
the uncertainties in the measured galaxy peculiar velocities (see below).  
Either trajectory from Fig.~1 can be used in fitting the data to determine 
$R$ and h.  For definiteness in our fitting procedure, we use the infall 
trajectory which corresponds to $\Omega =1$ and assume its validity for the 
general cosmological model (if, instead, we had used the $\Omega \rightarrow 
0$ trajectory, the change in $R$ would have been $\approx 3$\%).  To avoid 
confusion, note that in this approach Eq. (\ref{infall_trj}) cannot be used 
to find $h=h(t)$ (this equation gives $H =2/3t$, as it should for $\Omega = 
1$), but $h=h(t)$ has to be taken to match the underlying cosmology.

In applying spherical infall to the Local Group there is a complication.
Interior to the turnaround surface there are two large galaxies of roughly 
equal mass: the Milky Way and Andromeda (M31), M31 being somewhat larger.  
In what follows we adopt, following Peebles (1996), for the ratio of their 
masses, $M_{\rm MW}=0.7 M_{\rm A}$.  The derived mass and the turnaround 
radius of the LG will be a consequence of the gravitational pull of both 
galaxies.  We assume that those tracer galaxies that are sufficiently far 
away are infalling to the common center of mass of the Milky Way/M31 system. 

We utilize $\chi^2$ modeling of the infall trajectory, Eq. 
(\ref{infall_trj}), for the whole sample of galaxies in the local volume 
(excluding those satellites close to the Milky Way and Andromeda).  This 
may result in a more statistically robust determination of $R$ compared 
to the direct ``weighing'' at some selected positions through the use of 
equations (3) and (\ref{sol2}).

In the spherical infall modeling of the Local Group defined in this way, 
M31 is used only to determine the LG center of mass and does not appear 
explicitly on the infall diagram.  However, in the description of the 
M31/MW system another approximation can be made, namely that of two, 
isolated, mutually gravitating, compact bodies.  Using this approximation, 
the total mass of the system can be derived.  This is the so-called 
``timing argument'' of Kahn \& Woltjer (1959).

\section{M31 timing}

Let us assume that, at present, the relative motion of the Galaxy and M31 
can be described as motion of two, mutually gravitating, compact bodies
with masses $m_1$ and $m_2$.  The conserved total energy of the system is
\begin{equation}
E_{\rm tot} = \frac{m_1 m_2}{m_1+m_2}\left[\frac{1}{2}V^2 - 
\frac{GM}{d} \right] \, ,
\label{timing1}
\end{equation}
where $M \equiv m_1+m_2$, $V$ is velocity of M31 as seen from the Galaxy,
and $d$ is the distance between them. This gives the equation of motion
which is identical to Eq. (1) but with E being replaced by 
$\tilde{E} \equiv E_{\rm tot}M/m_1m_2$.
Since $\tilde{E}$ is unknown (as is E), this variable can be replaced by 
another variable, $t$, using solutions, Eqs. (2).
In this way Eqs. (3) - (4) are reproduced with the coordinates relative to 
the center of mass, ${r,v}$, being replaced by coordinates relative to the 
Galaxy, ${d,V}$, and $M$ is the total mass of the MW/M31 system.  Using the 
observed values of $d,V,$ and $t$, $M = M(d,V,t)$ can be calculated.  As we 
shall see, the mass calculated in this way exceeds the predictions of the 
spherical infall by about a factor of two.  This is a serious internal 
contradiction in our modeling of LG dynamics and it has to be addressed.

There are several caveats in the M31 timing argument.  $\tilde{E}$ can 
be related to $t$ only if the equation of motion, Eq. (\ref{timing1}), 
is valid at all times.  But this is not generally true.  The dominant
galaxies started to grow from small initial fluctuations which were
{\it extended} in space.  Their masses were not constant, but grew because 
of infall and the merging of smaller, progenitor galaxies.\footnote{These 
objections do not apply to the spherical infall model as long as spherical 
symmetry holds.} 
The ``timing'' approach can be modified assuming that $m_i = m_i(t)$,
with the growth rate taken from the spherical infall model for each 
galaxy halo separately, and assuming that the resulting motion of galaxies 
can be described as the motion of point particles on an unperturbed
cosmological background.  However, there is a caveat to this approach too.  
The background cannot be considered as unperturbed, especially at those late 
stages when large extended halos grow at the expense of surrounding material.

The accuracy of the timing argument with respect to the neglect of finite 
size effects can be tested using the results presented by Peebles et al. 
(1989), where LG formation was modeled using N-body simulations.  The 
results of Peebles et al. (1989) can be considered as a generalization 
of M31 timing which accounts for the finite sizes and for the finite 
perturbations of the cosmological background.  Results were presented 
there for two values of the present age, $t= 7.8$ Gyr and $t=13.8$ Gyr.  
For $t= 7.8$ Gyr, the mass of the Milky Way/M31 system derived from the 
timing argument was $7.6\times 10^{12} M_\odot$, while the LG mass inside 
the turnaround surface measured in the N-body experiment was $4.9\times 
10^{12} M_\odot$.  For $t=13.8$ Gyr, the timing argument gives $4.6\times 
10^{12} M_\odot$, while the mass inside the turnaround surface was found 
to be $3.5\times 10^{12} M_\odot$ in the N-body simulation.  The agreement 
is not very good, with M31 timing overestimating the mass inside the 
turnaround surface.  The calculations of Peebles et al. (1989) were 
restricted to an $\Omega_0 =1$ universe; we are unaware of analogous 
results for an open universe model.  In contrast, the predictions of 
the spherical infall model are valid irrespective 
of the value of $\Omega_0$. 

Last but not least, from our point of view the most serious drawback of the 
M31 timing is that it uses a single observation.  Since unknown peculiar 
velocities are inevitably present in the motion of any galaxy, the results 
derived from just one data point can deviate significantly from the mean.  
In contrast, when the motion of a large set of ``test particles'' is 
considered, which can be done in spherical infall modeling, the peculiar 
velocities may average out.  For this reason we prefer the spherical infall 
approach to the M31 timing argument.  Unfortunately, not many galaxies are 
available for modeling the motion near the turnaround surface and it would 
be inappropriate to ignore the information provided by the motion of M31.
For these reasons we adopt the following procedure.

We assume that all galaxies belonging to the Local Group have started in 
a common initial environment.  Therefore, we can try to relate $\tilde{E}$ 
in Eq. (\ref{timing1}) to $E$ which enters the equation of motion for the 
other, outlying satellites of the Local Group, Eq. (\ref{eqm}).  (Each 
satellite has its own value of E, determined by its initial position.)  
In this way it is possible to put M31 on the infall diagram along with 
the rest of the galaxies.  Assuming two initial small overdensities 
(one the seed of the Galaxy, the other the seed of M31) on an otherwise 
unperturbed initial Hubble flow, we find that the data points (r,v), 
where r and v are the distance and velocity of the Local Group satellites 
as seen from the center of mass, along with the data point (d,V), 
corresponding to M31 as measured from the Galaxy, all belong to one and 
the same infall trajectory to a very good approximation.  This procedure, 
which would overestimate the mass of the Local Group if there were an 
extra mass excess (say, unseen matter residing in the center of mass), 
will be our basic approach to bounding mass of the Local Group from above.

\section{Infall in the Local Group.}

Our basic assumption is that those tracer galaxies sufficiently far away 
are infalling to the common center of mass of the MW/M31 system.  In addition, 
we have to make an assumption about the relative velocity of the Milky Way.
We assume 
that, relative to the sample of sufficiently close galaxies, $r < D$, the 
Milky Way also infalls to the common center of mass.  However, the Local 
Group as a whole may be moving with respect to the more distant galaxies.  
To account for this possible local deviation from the Hubble flow, we 
determine from our fitting procedure the Milky Way velocity relative to 
the sample of galaxies at $r > D$, as well as the distance D itself.  We 
utilize the $\chi^2$ approach to modeling the infall trajectory; i.e. we 
look for the minimum of the sum
\begin{equation}
\sum_i \left[ v(i) + \vec{v}_{\rm MW} \cdot \vec{e}(i) - 
v(r) \, \vec{e}_{\rm cm} \cdot \vec{e}(i) \right] ^2 \, \, ,
\label{chi2}
\end{equation}
where $v(i)$ is the measured velocity of $i$-th galaxy, $\vec{e}(i)$ 
is the unit vector pointing from the center of mass to the direction 
of that galaxy, 
%(with $l$ and $b$ being angles in Galactic coordinates, see below), 
$\vec{e}_{\rm cm}$ is the unit vector pointing from the Galaxy to the 
center of mass (i.e., according to our assumption, in the direction 
of M31), $v(r)$ is the infall trajectory, Eqs. (5), (\ref{infall_trj}), and 
$\vec{v}_{\rm MW}$ is the Milky Way velocity.  Relative to a set of 
galaxies with $r(i) < D$, the Milky Way velocity is assumed to be 
$v_{\rm MW} = v_{\rm M31} (1+M_{\rm MW}/M_{\rm M31})^{-1}$.  M31 
itself is included in this fit, as explained in the previous Section.  
We do this modeling in \S 4.1.  For comparison, in \S 4.2, we apply 
equations (3) - (4) for the direct ``weighing" of the Local Group, 
using the subset of galaxies which lie closest to the turnaround sphere.

We don't expect the spherical approximation to be accurate for nearby
galaxies with distances from the center of mass of the LG comparable 
to their distances from one of the dominant galaxies.  We exclude from 
the fit Eq. (\ref{chi2}) the close satellites of either the Milky Way 
or M31.  On the other hand, radial infall in the direction of one of 
the dominant galaxies may be a good approximation for close satellites 
(so that the satellite does not ``feel'' the other heavy galaxy).  Most 
such satellites are inside the first caustic of the infall model, where 
their trajectories are very model dependent.  Luckily, there are two 
satellites that might be just outside the first caustic.  Using Eq. 
(\ref{sol2}), in \S 4.3 we derive the mass interior to these 
satellite galaxies under the assumption of radial infall to one of the 
dominant galaxies. 

\subsection{Modeling the infall trajectory.}

We included in our fit to Eq. (\ref{chi2}) all galaxies in Karachentsev 
\& Makarov (1996) at distances less than 3~Mpc from the center of mass 
of the Milky Way/M31 system; we excluded those whose distances from either 
the Milky Way or M31 are smaller than 0.3 Mpc.  The resulting best fit 
trajectory, as well as the data points recalculated relative to the center 
of mass are shown in Fig.~2.  Data points are connected by the dotted line 
in order of increasing $r$; the first ten galaxies are listed in Table I.

The parameters of the best fit trajectory are: $R=0.93$ Mpc, $h=0.69$ 
and $D =1.5$ Mpc. The formal $\Delta \chi^2 = 2.3 $ confidence contour
in the $R$, $h$ plane is shown by the solid line in Fig.~3.  { After 
we had completed the analysis presented here a new data set compiled by 
Mateo (1998) became available.  According to Mateo, the distance of Leo A, 
the ``low" velocity point at $r = 1.6$~Mpc in Fig.\ 2 (see also Fig.\ A8), 
should be smaller by a factor of two, placing it right on our best fit 
trajectory.  We have repeated the analysis described here using Mateo's 
data, finding $R = 0.9$ Mpc and $h = 0.6$.}
 
{\plotone{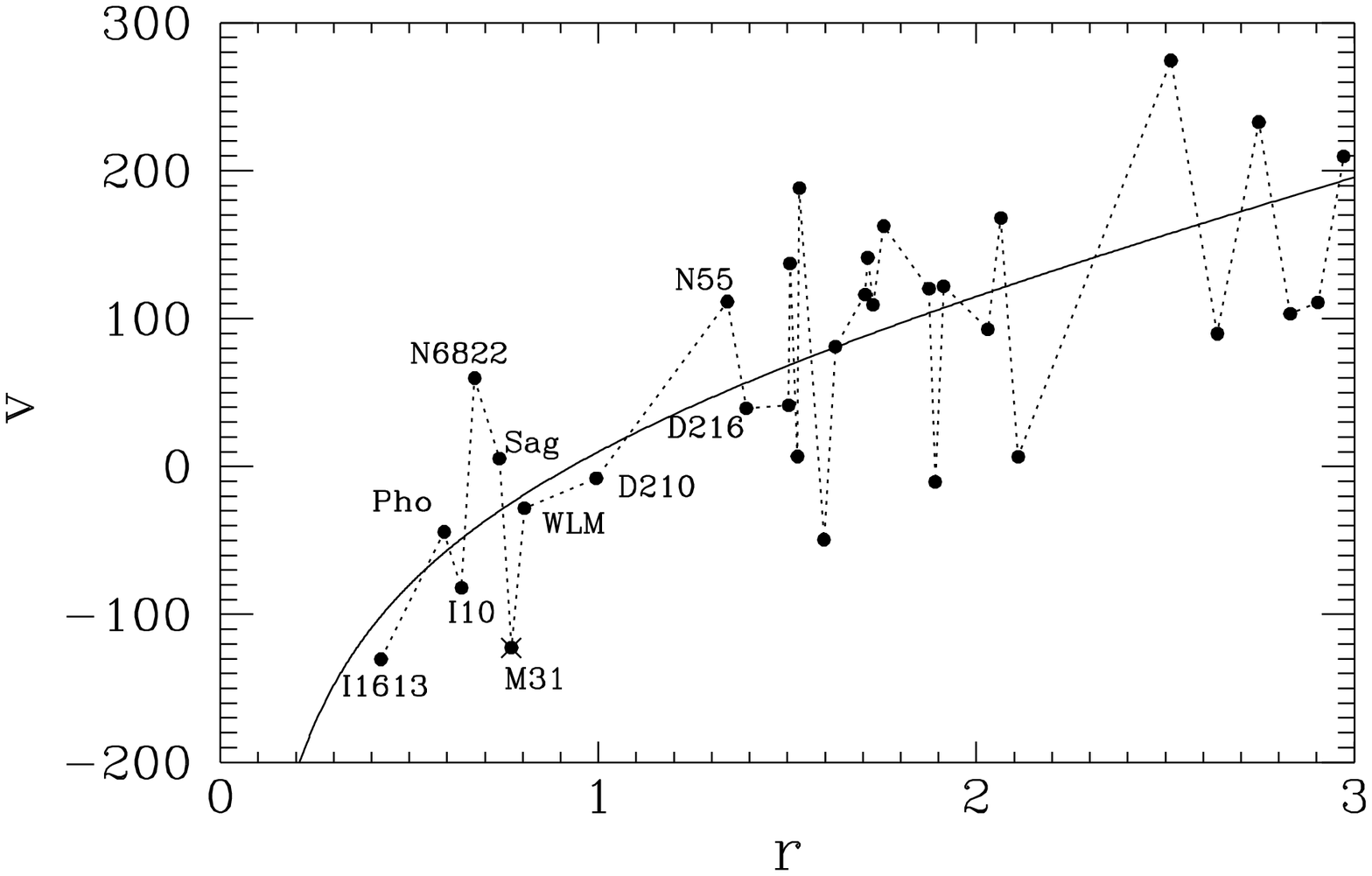}
\figcaption{
\baselineskip0.08cm
Phase space for the sample of nearby galaxies deriveded under the 
assumption of infall to the center of mass of the Milky Way -- M31 
system.  The distance ($r$) from the center of mass is in Mpc, the 
infall velocity (v) is in km/s.  For M31, $r$ and v are the distance 
and the velocity relative to the Galaxy.  The best fit infall trajectory, 
corresponding to $R=0.93$ Mpc and $h=0.69$, is shown by the solid line. 
\label{rv}}
}

\begin{table}
\caption{
\baselineskip0.08cm
The sample of Local Group galaxies.  $l$ and $b$ are 
galactic coordinates; $r_g$ and $v_g$ are galactocentric
distance (Mpc) and velocity ($km/s$); $r$ (Mpc) and 
$v$ ($km/s$) are the distance and velocity with respect 
to the center of mass of the Milky Way/M31 system.}
\vspace{0.3cm}
\begin{tabular}{cccccccc}
& Name & $l$ & $b$ & $r_g$ & $v_g$ & $r$ & $v$ \\
\tableline
0 & M31   & 121.2 & -21.6 & 0.77 & -123 &    &     \\ 
1 & IC1613 & 130 & -60.6 & 0.66 & -152 & 0.425 & -130 \\ 
2 & Phoenix & 272 & -69 & 0.42 & -33.3 & 0.592 & -44.2 \\ 
3 & IC10 & 119 & -3.3 & 1.04 & -146 & 0.638 & -81.9 \\ 
4 & N6822 & 25.3 & -18.4 & 0.52 & 43.4 & 0.673 & 59.8 \\ 
5 & Sagitt & 21.1 & -16.3 & 0.57 & 8.11 & 0.738 & 5.46 \\ 
6 & WLM & 75.7 & -73.6 & 0.95 & -62.6 & 0.804 & -28.1 \\ 
7 & DDO210 & 34.1 & -31.4 & 1 & -23.4 & 0.995 & -7.94 \\ 
8 & N55 & 333 & -75.7 & 1.34 & 94.2 & 1.34 & 112 \\ 
9 & DDO216 & 94.8 & -43.6 & 1.75 & -21.1 & 1.39 & 39.5 \\ 
\end{tabular}
\label{I}
\end{table}

{\plotone{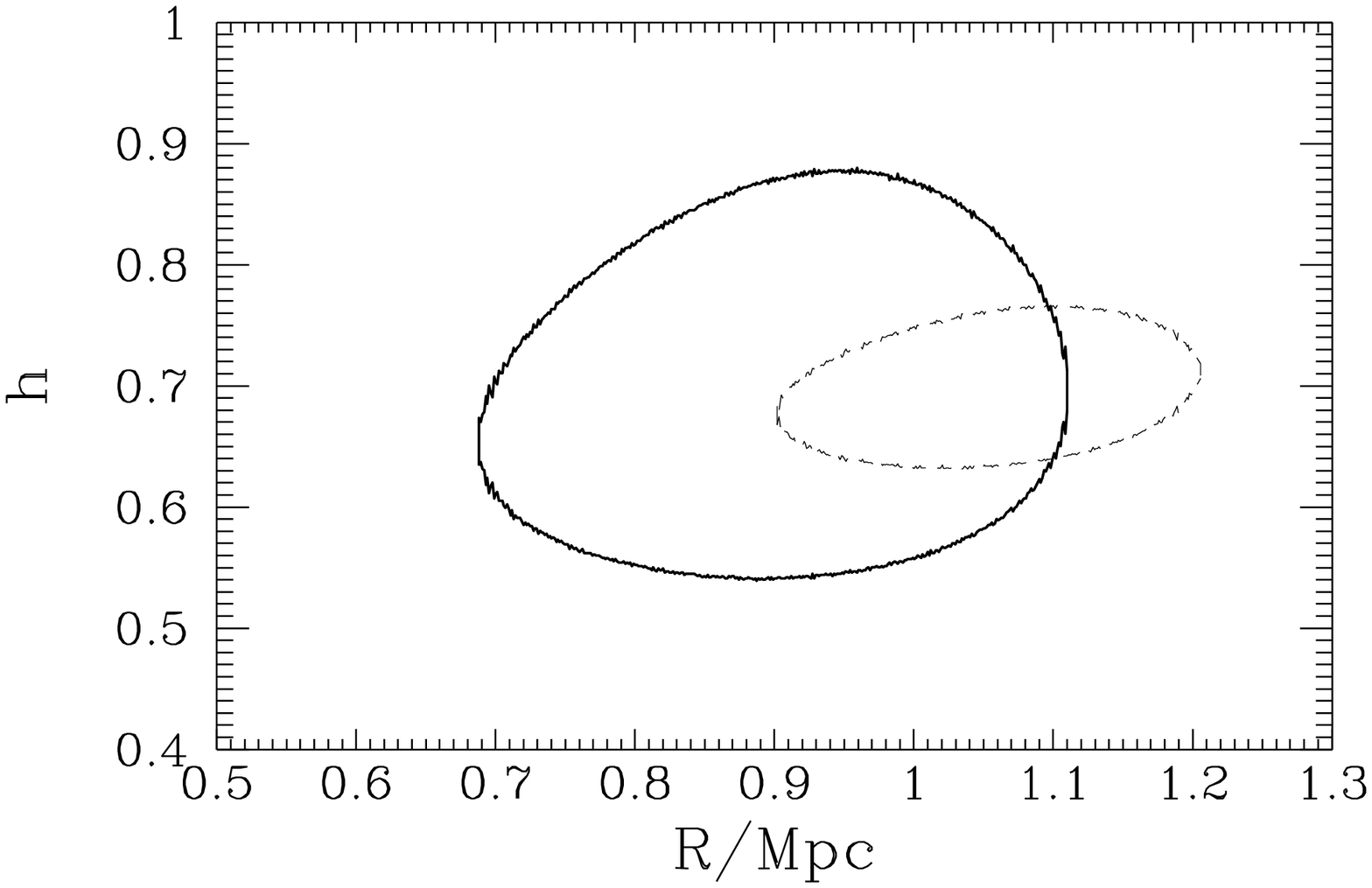}
\figcaption{
\baselineskip0.08cm
The $\Delta \chi^2 = 2.3 $ confidence contour in the Hubble 
parameter ($h$) -- LG turnaround radius ($R$) plane is shown 
by the solid line.  The dotted line corresponds to the case 
when galaxies forming subclusters were removed from the fit;
see the Appendix for details. 
\label{Fig:conf}}
}
\vspace{0.3cm}

It was assumed that, with respect to galaxies in the volume whose scale 
is $r < D$, the Milky Way infalls to the common center of mass, while 
we fitted the Galaxy velocity with respect to galaxies in the volume 
$D < r < 3$ Mpc.  This allows us to take into account the possible motion 
of the Local Group as a whole.  If we were to neglect the Local Group 
velocity, i.e., if $D = \infty$ were assumed, the turnaround radius would 
be smaller, the value of the Hubble constant would be larger, and the 
velocity dispersion would increase.  Our fit is made using the value 
of the Milky Way velocity found with respect to the whole set of galaxies 
in this volume, not with the velocity being allowed to vary with $r$. 
It is, however, instructive to study the trend of the Milky Way velocity 
derived with respect to the galaxies in volumes of varying depth.  In this 
way we can test the self-consistency of the assumption of infall to the 
center of mass of the Milky Way/M31 system in the volume $r < D=1.5$ Mpc.  
We defer a discussion of this trend to Appendix A.

We may conclude that $r = D \approx$ 1.5 Mpc is a boundary of the Local
Group.
Note that despite the possibility that many galaxies in the sample 
at $r > D$ may be forming gravitationally bound groups of their own, 
which would induce additional peculiar velocities, the velocity 
dispersion turns out to be very small, $\sigma_v = 67$ km s$^{-1}$ 
{ (using Mateo's (1998) data set, $\sigma_v = 50$ km s$^{-1}$)}. 
This value, which is significantly smaller than what might be expected 
in the standard ($\Omega_0 = 1$) CDM, $\sigma_v \sim 500$ km s$^{-1}$ 
(\cite{GB94}), may be a signature of low density ($\Omega_0 < 1$).  

Pairwise velocities of gravitationally bound pairs of galaxies or, more 
generally, the local infall and virialized components of velocites of 
galaxies which form remote clusters, make significant contributions to 
this velocity dispersion.  One might expect that the picture of infall 
onto the Local Group will be ``cleaner'' if close pairs of galaxies are 
removed and only relatively isolated galaxies are selected for the fit.
This procedure and the results are discussed in Appendix A.  The resulting 
confidence contour in the ${R, h}$ plane is shown by the dashed line in 
Fig.~3; the best fit values are: $R=1.07$ Mpc, $h=0.71$.  The best fit 
turnaround radius obtained when all galaxies are included is $R=0.93$ Mpc; 
in the case when subclusters were removed, $R=1.07$ Mpc.  These are within 
``1 $\sigma$'' of each other.  While the procedure with the subclusters 
removed would seem to be more promising, supported by the significantly 
lower value of the resulting velocity dispersion, at present we choose to 
take as our basic estimate of turnaround radius the average of $R_{\rm ta}$ 
obtained in both procedures; this gives\footnote{The error estimate has 
been slightly biased towards a larger value of $R$ to make the errors 
symmetric.}
\begin{equation}
R_{\rm ta} = (1.0 \pm 0.2) \,\, {\rm Mpc} \,\, . 
\label{Rta}
\end{equation}
Using equations (\ref{sol}) we can relate the mass inside the turnaround 
sphere to the turnaround radius as
\begin{equation}
M_{\rm ta} = \frac{\pi^2 R^3}{8G t^2} = 2.74 \times 10^{12} M_\odot \, 
\frac{R^3}{t_{10}^2} \, ,
\label{mta}
\end{equation}
where $R$ is in Mpc and $t_{10} = t/10$ Gyr. Using Eq. (\ref{Rta})
and accounting for the propagation of errors, we obtain 
$M_{\rm ta} = (3.1 \pm 1.6) \times t_{10}^{-2}\, 10^{12} M_\odot$. 
Taking 7/17 of this to be the Milky Way fraction, we find $M_{\rm ta}(MW) 
= (1.3 \pm 0.7) \times t_{10}^{-2}\, 10^{12} M_\odot$.  The mass that is 
just outside of the first caustic should be even smaller.  To find this 
latter mass in the $\epsilon =1$, $\Omega_0 =1$ model, these numbers 
should be multiplied by $0.7$, giving

\begin{equation}
M_{\rm 1}(MW) = (0.88 \pm 0.47) \times t_{10}^{-2}\, \times 10^{12} 
M_\odot \, .
\label{mmw}
\end{equation}

We adopt these as our basic estimates of the mass of the Local Group 
and of the Milky Way.  The further discussion in subsequent \S 4.2 
and \S 4.3, is mainly for comparison.

\subsection{Direct weighing of the LG using tracers of infall near 
the turnaround sphere.}

Among those galaxies near the turnaround sphere, the galaxies IC 1613, 
WLM, DDO 210 and DDO 216, are closest relative to our best fit infall 
trajectory (see Figs.~2. and A1).  On the other hand, N6822 and M31 
appear to be outliers.  It is, therefore, interesting to analyse IC 1613, 
WLM, DDO 210 and DDO 216 separately,\footnote{Phoenix and IC 10 are closer 
to one of the dominant galaxies than to the center of mass (see Tables I 
and II), therefore we analyse them with the assumption of infall to one 
of the large galaxies; see below.}
using Eqs. (\ref{sol2}) to ``weigh'' the Local Group directly, without 
any assumptions about the shape of the initial inhomegeneity or the 
cosmological model.  The unknown cosmological parameter of \S 4.1,
the Hubble constant $h$, is replaced here by the cosmic time, $t$.  
Since the relation $h = h(t)$ is model dependent and unknown at present,
this ``direct weighing'' may be of some interest.

We plot the total mass interior to the tracer galaxies versus cosmic 
time in Figure~4.  We warn the consumer that conclusions derived from 
analysis of only a handful of galaxies should be taken cautiously.  { 
Although Mateo (1998) estimates typical distance errors in the 5 -- 10\%
range, some galaxies in his sample have had their distance estimates vary
by a factor of two.}  To illustrate the effect of errors in the distance 
estimates, we draw with dotted lines in Figure~4, a~ $\pm 10$\% error 
``corridor'' around IC1613.

{\plotone{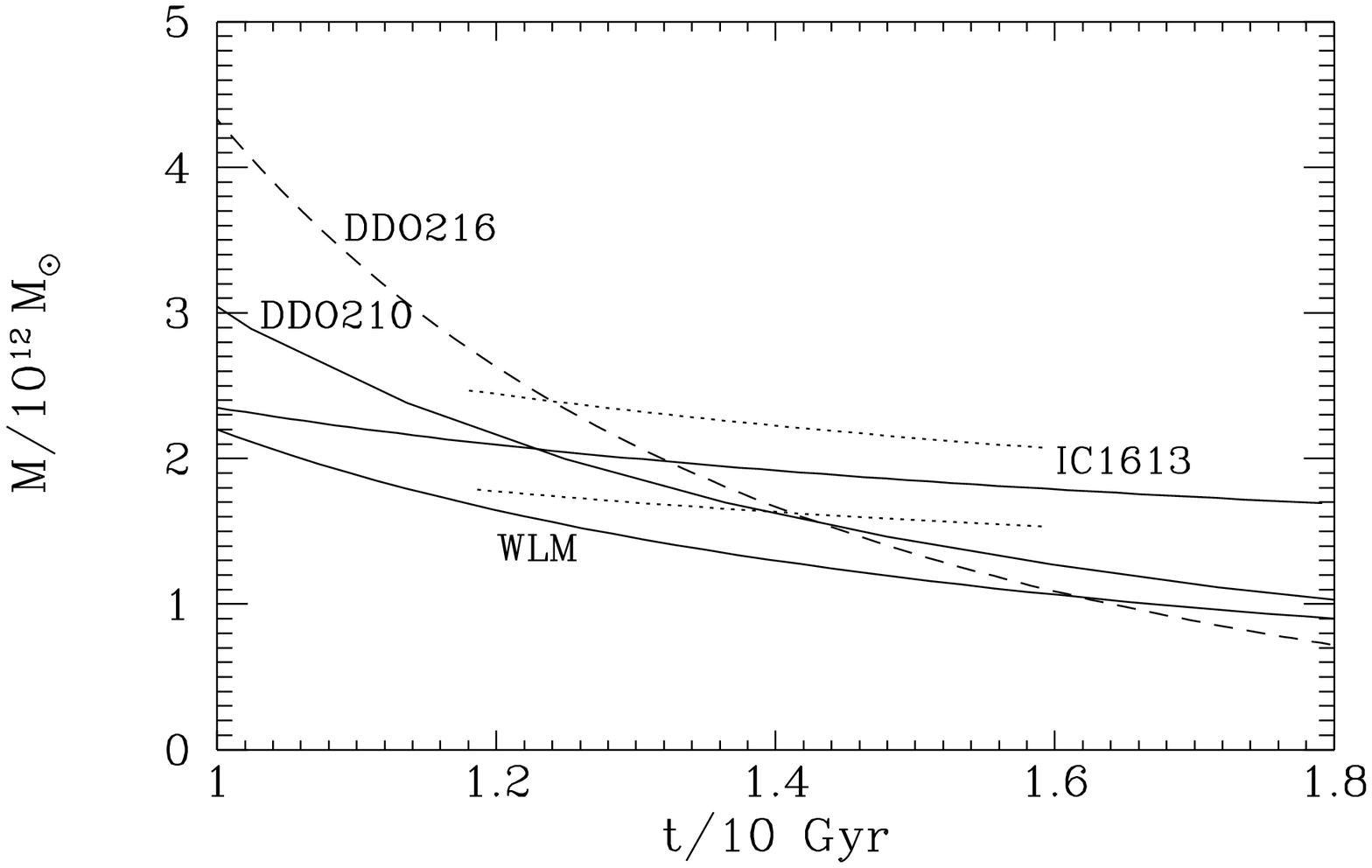}
\baselineskip0.08cm
\figcaption{Total mass interior to a tracer galaxy versus 
the age of the Universe (dashed and solid curves).  The dotted curves
represent the $\pm 10$\% error ``corridor'' around IC1613.
\label{4gal}}
}
\vspace{0.3cm}

IC 1613 is probably too close for the spherical infall model to be 
reliable.  In contrast, DDO 216 is outside the turnaround surface and 
is therefore still expanding with the Universe; for this reason we plot 
this galaxy with a dashed line in Figure~4.   The best tracer galaxies 
could be WLM and DDO 210.  DDO 210 is very close to the turnaround 
surface but is already inside it.  We find $M(DDO210) = 3.0 \times 
t_{10}^{-2}\, 10^{12} M_\odot$, in good agreement with the Local Group 
mass estimate from Eq. (\ref{mta}).  It appears that WLM may have fallen 
in even earlier, providing an even stronger constraint on the mass within 
the first caustic.  If we adopt 10 Gyr as a lower bound for the age of the 
Universe, we infer (see Fig.~4) $M(0.8\, {\rm Mpc}) < 2.2 \times 10^{12}M_
\odot$.  For the total mass associated with the Milky Way we have to take 
7/17 of this value, leaving the rest for M31; this gives $M_{\rm MW} \alt 
0.9 \times 10^{12} M_\odot $.  This is in agreement with our best fit estimate 
for the Milky Way fraction of the mass inside the first caustic (Eq. 11).  
[We recall that since $R_{\rm ta} \sim 1$~Mpc, the radius of the first caustic 
$R_{1} \sim 0.37 R_{\rm ta} \sim 0.4$~Mpc and consequently WLM is outside 
the first caustic, although not very far from it.]  Note that since the 
MW -- M31 distance is larger than twice the radius of the first caustic, 
there has been no mixing between material that infalls to the Milky Way 
and that which infalls to M31.  

Since DDO 216 is the galaxy farthest away from the center of mass,
(see Table~I), the total mass interior to its orbit should be the 
largest.  However, contrary to this expectation, the dashed line 
on Figure~4 falls below the solid lines for $t > 13 - 16$ 
Gyr.  This should not occur in the absence of observational errors 
(but, note the effect of 10\% errors on the IC 1613 curve), provided 
that spherical infall is an adequate approximation.  If so, and in 
the absence of errors, we could derive an upper bound to the age of 
the Universe, $t \alt 13 - 16$ Gyr.  Interestingly, this suggestive 
bound limits the age of the Universe from above, while the more 
classical method, based on the ages of the oldest stars, provides 
bounds from below. 

\subsection{Infall to one of the dominant galaxies}

In this subsection we present the results of an analysis of Phoenix and IC10, 
nearby satellite dwarf galaxies that are closer to one of the dominant galaxies 
than to the LG center of mass, and which lie in the hemisphere opposite the 
other dominant galaxy.  At the same time, they are sufficiently far away to 
be outside the first caustic.  The approximation of direct infall to the 
closest dominant galaxy might not be unreasonable for these dwarfs.  The 
relevant parameters are listed in Table~\ref{II}. 

\begin{table}
\caption{
%\footnotesize
\baselineskip0.08cm
Same as in Table I, except now $r$ and $v$ are the distances 
and velocities with respect to the center of the Milky Way for 
Phoenix, and with respect to the center of M31 for IC10.}
\vspace{0.3cm}
\begin{tabular}{ccccccc}
Name & $l$ & $b$ & $r_g$ & $v_g$ & $r$ & $v$ \\
 \tableline
Phoenix & 272.2 & -69.0 & 0.42 & -33  & 0.42 & -33  \\
IC10    & 119.0 & -3.3  & 1.04 & -146 & 0.39 & -38  \\
\end{tabular}
\label{II}
\end{table}

We have assumed that Phoenix infalls to the Milky Way and that IC10 infalls 
to M31. The mass interior to these satellite galaxies is shown in Figure~5 
along with a 10\% ``error corridor" for the distance estimate.  Although 
this results in a smaller inferred mass than our best fit estimate 
(Eq. \ref{mmw}), it is within the estimated ``$1 \sigma$'' error.

{\plotone{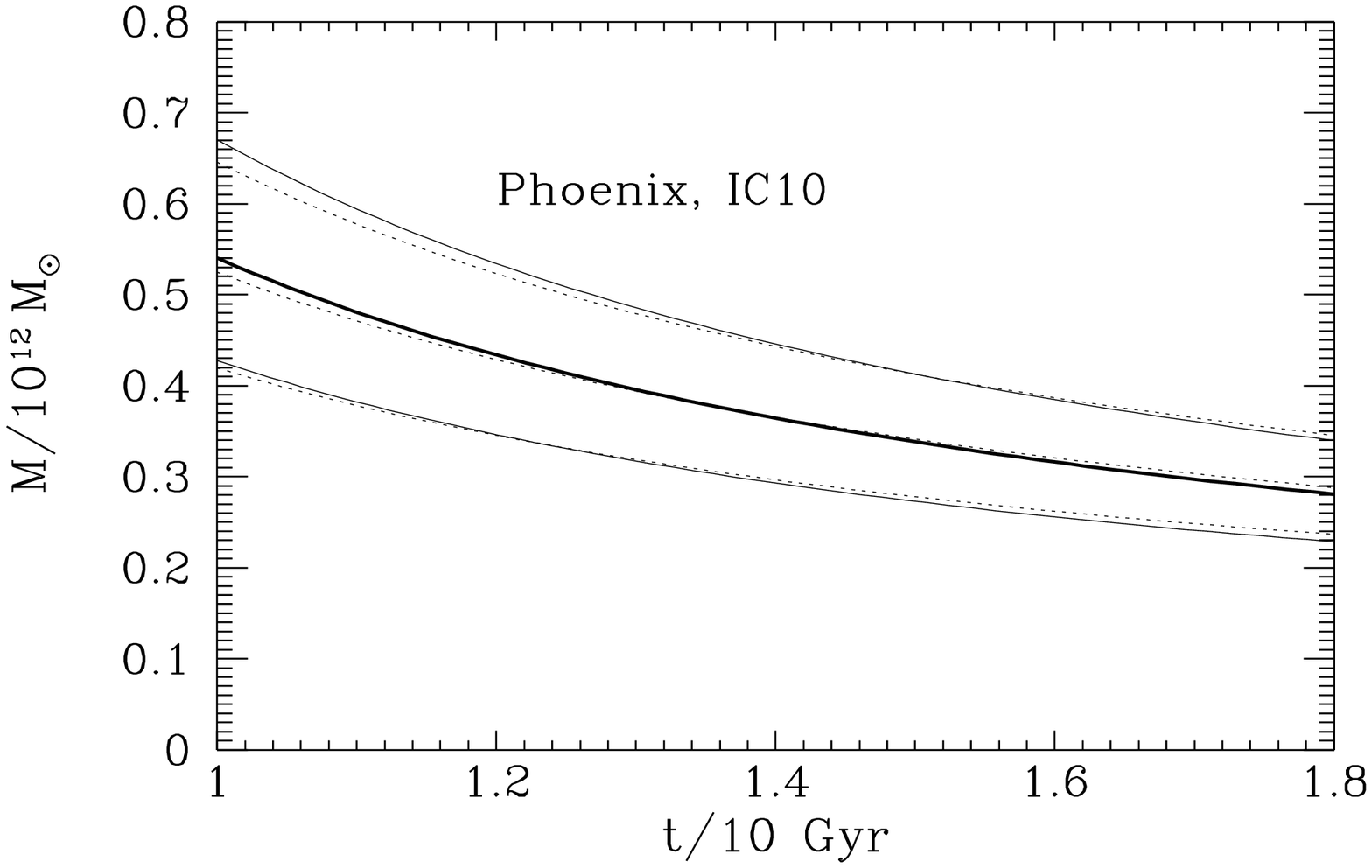}
\figcaption{
\baselineskip0.08cm
Same as Figure~4 for Phoenix (solid curves) and IC10 (dotted curves).
\label{ombfig3}}
}
\vspace{0.3cm}

\section{Primordial baryon fraction and $\Omega_0$}

Using equation (\ref{mmw}) along with the MACHO results and the standard 
formulae for the propagation of errors we obtain a lower bound for 
the universal primordial baryon fraction $f_{\rm B} \agt M_{\rm B} 
(50~{\rm kpc})/M_{\rm MW} = 0.29^{+0.18}_{-0.15}\, t_{10}^2$.  This 
can be rewritten as 
\begin{equation}
\Omega_{\rm B}  \agt \, 0.29^{+0.18}_{-0.15}\, t_{10}^2 \,\Omega_0 \, .
\label{omb}
\end{equation}
Introducing the baryon-to-photon ratio $\eta = n_{\rm B}/n_\gamma$ so 
that $\Omega_{\rm B} h^2 = \eta_{10}/273$, where $\eta_{10} \equiv 
\eta/10^{-10}$, we may rewrite this equation to derive an upper bound 
on the total matter density, $\Omega_0 h^2 t_{10}^2 \alt 3.66 \times 
10^{-3} \eta_{10} /f_B $.  In the absence of a cosmological
constant the age and Hubble parameter are related by
\begin{equation}
t_{0}H_{0}=
\int_0^1 \left(1-\Omega_0+\Omega_0 x^{-1}\right)^{-1/2}dx, 
\label{tH}
\end{equation}
where 
$H_{0}^{-1} =9.78 h^{-1}$ Gyr, so that  
\begin{equation}
\Omega_{0}\,(H_{0}t_{0})^2  \alt \, 0.020^{+0.010}_{-0.013}\, \eta_{10}\, .
\label{om0}
\end{equation}
Similarly, $t_{0}H_{0}$ can be found as a
function of $\Omega_{0}$ for any given value of 
$\lambda_0 \equiv \Lambda/3H_0^2$.
The left-hand side of equation~(\ref{om0}) is plotted versus $\Omega_{0}$ in 
Figure~6 for two cases, $\lambda_0 = 0$ and $\Omega_{0}+ \lambda_0 = 1$.

{\plotone{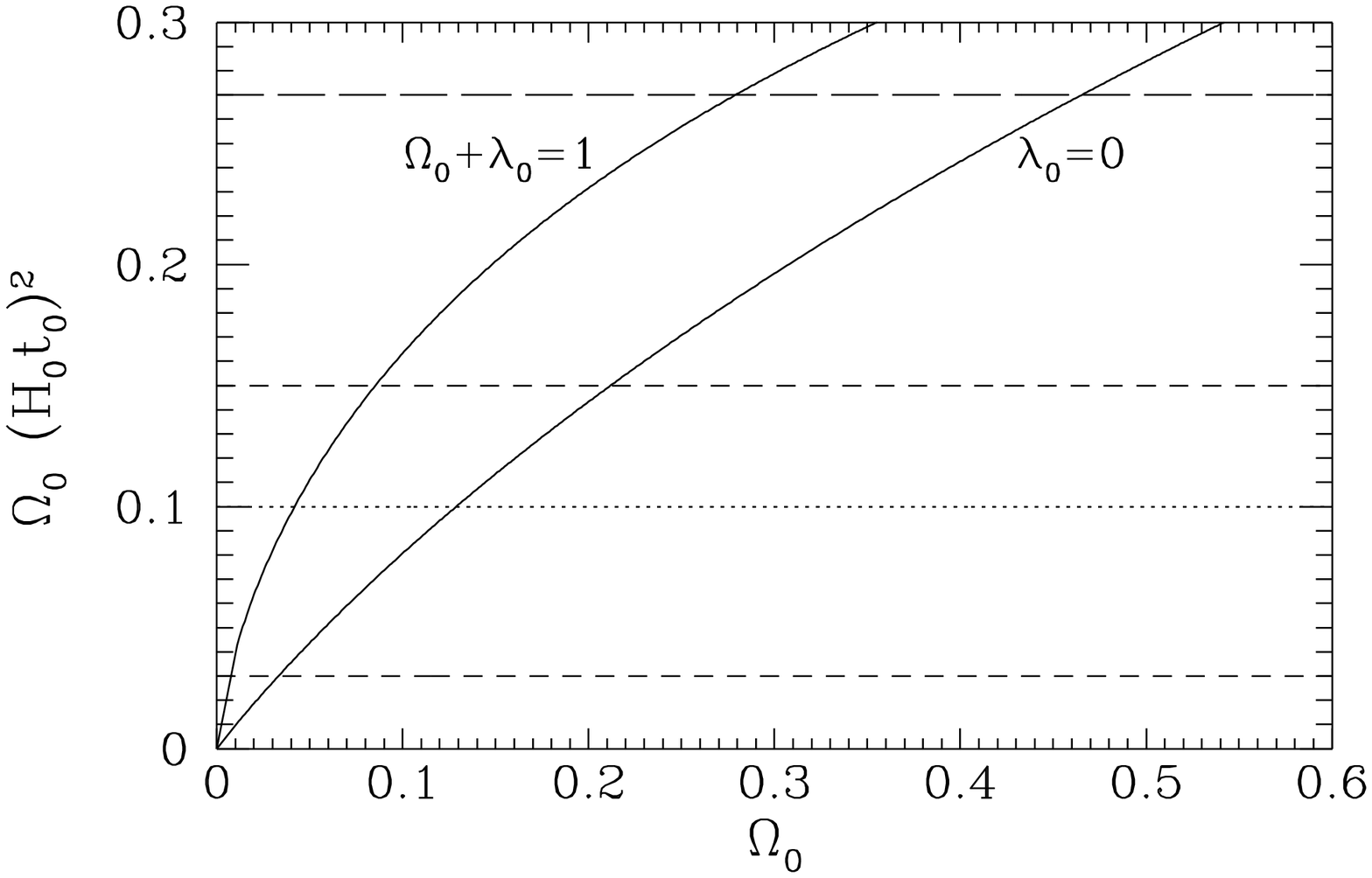}
\figcaption{
\baselineskip0.08cm
$\Omega_{0}(H_{0}t_{0})^2$ versus $\Omega_{0}$ (solid curves) along
with our bound, Eq. (\ref{om0}). The dotted and dashed lines are for 
$\eta_{10} = 5.1 \pm 0.3$ respectively, while the long-dashed line 
is for the upper bound in Eq. (\ref{om0}) and $\eta_{10} < 9.0$.
\label{ombfig4}}
}
\vspace{0.3cm}

Using SBBN to bound the right-hand side of Eq.~(\ref{om0}), we may adopt 
for the nucleon-to-photon ratio the Burles \& Tytler (1998) deuterium-driven 
estimate $\eta_{10} = 5.1 \pm 0.3$.  From Figure~6 this leads to the bound, 
$\Omega_0 = 0.13^{+0.08}_{-0.10}$ for $\lambda_0 = 0$, or $\Omega_0 = 
0.04^{+0.04}_{-0.03}$ for $\Omega_{0}+ \lambda_0 = 1$.  As a conservative 
upper bound we may adopt $\Omega_0 \alt 0.2$.  Note that, for an extreme 
upper bound to the nucleon-to-photon ratio derived from the lithium abundance 
alone (\cite{bbn}), $\eta_{10} < 9.0$, we constrain  $\Omega_0 < 0.47$ for 
$\lambda_0 = 0$ and $\Omega_0 < 0.28$ for $\Omega_{0}+ \lambda_0 = 1$ (see 
Figure~6).  Note, however, that if the vacuum energy is cosmologically 
relevant, the equations of motion near the turnaround radius will be 
modified {\it independent} of the value of R.  Therefore, in this case 
Eq. (10) is modified and, correspondingly, so too will be our bounds for 
the baryon fraction.  However, the effect of a cosmological constant is 
to {\it decrease} $M_{\rm ta}$ at a given R and t (i.e. to {\it increase} 
$f_B$).  Therefore, our bounds quoted here are conservative for the case 
of non-zero vacuum energy.

This preference for a low value of $\Omega_0$ 
was already noted in Peebles et al. (1989) and Peebles (1995, 1996) 
based on the motions of the LG galaxies, and is consistent with recent 
observations on larger scales (see e.g. \cite{bld}; \cite{Kashl}; 
\cite{xray2}; \cite{will}; \cite{perl}).

It is interesting to compare our constraint on the universal baryon 
fraction, $f_B \agt 0.14~t_{10}^2$, with the baryon fraction derived 
from X-ray clusters, $f_B\, h^{3/2} = (1.0 \pm 0.1)(1+h^{3/2}/5.5)/15$ 
(\cite{xray2}; \cite{xray1}).  For $t_{10} \agt 1$, both constraints 
are shown in Figure~7.  Although there is reasonable agreement for 
$t_{0} = 10$ Gyr, larger ages provide a hint of a discrepancy.  Such 
disagreement may point towards more dark baryons in clusters (e.g., 
intergalactic MACHOs (\cite{AG95})) than is revealed by the X-rays, 
or fewer LG baryons (non-baryonic halo dark matter).  

{\plotone{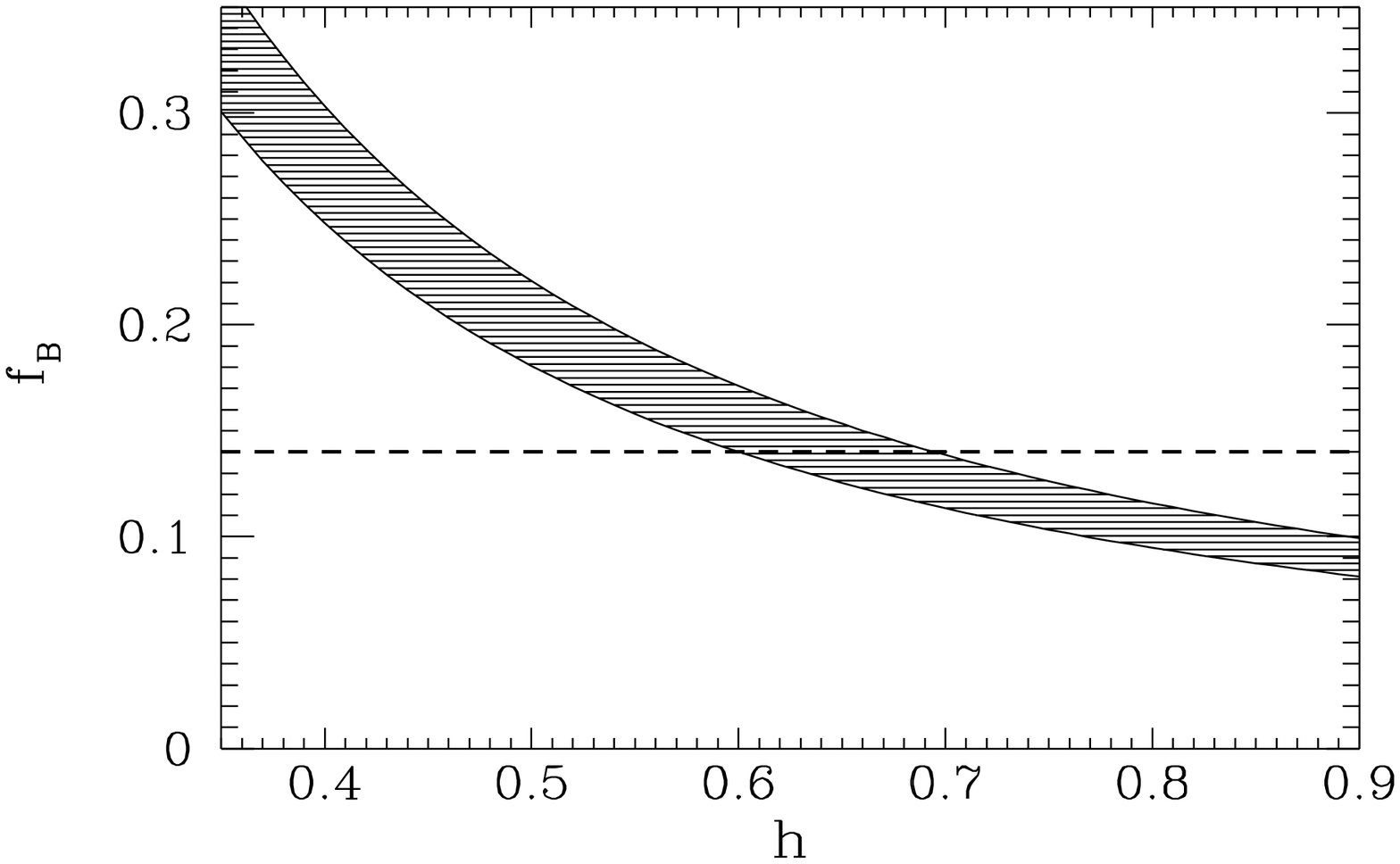}
\figcaption{
\baselineskip0.08cm
The baryon fraction -- Hubble parameter plane.  
The crosshatched region shows the estimate of the baryon fraction 
derived from X-ray clusters.  The dashed line is our LG constraint, 
$f_B > 0.14$ (for $t_{0} > 10 $ Gyr). 
\label{ombfig5}}
}
\vspace{0.3cm}
{ Alternatively, we may use our MW mass estimate (eq. 11) of 
$M_{1}(MW) \approx 8.8~t_{10}^{-2} \times 10^{11}M_{\odot}$, along with
the X-ray cluster baryon fraction estimate of $f_B \approx 1/15h^{3/2}$,
to ``predict" the MW baryon mass
\begin{equation}
M_{\rm B}(MW) \approx 5.9~h^{-3/2}t_{10}^{-2} \times 10^{10}M_{\odot}.
\label{B}
\end{equation}
For $t_{0} \approx 14$~Gyr and $h \approx 0.65$ (\cite{sn1a}), this estimate 
suggests a MW baryon mass, $M_{\rm B}(MW) \approx 6 \times 10^{10}M_{\odot}$, 
considerably smaller than the MACHO value, but consistent with the MW 
disk mass (\cite{fhp}); perhaps the bulk of the MACHO events are not 
caused by baryonic dark matter.}

\section{Conclusions}

There remain several uncertainties in our LG baryon fraction estimate.  
One possibility which would weaken or even eliminate our constraint is 
if some of the observed microlensing events towards the LMC were due 
to an intervening satellite galaxy between us and the LMC, or due to 
debris in the LMC tidal tail (\cite{z96}; \cite{z97}).  However, the 
MACHO collaboration concluded (\cite{fg}) that if the lenses were in 
a foreground galaxy, it must be a particularly dark galaxy; see also 
(\cite{AG97}).  Moreover, the first observation of a microlensing event 
in the direction of the Small Magellanic Cloud (SMC) (\cite{smc}), 
implies an optical depth in this direction roughly equal to that in the 
direction of the LMC.  This makes it unlikely that a dwarf galaxy or a 
stellar stream between us and the LMC is responsible simultaneously for 
the observed microlensing towards the LMC and the SMC (\cite{fg}; 
\cite{AG97}).  Recently, however, Gates et al. (1997) found Galactic 
models which explain the current microlensing data by a dark extension 
of the thick disk, reducing the MACHO fraction.  It is to be anticipated 
that as more microlensing data are accumulated, these uncertainties will 
be resolved.  

We note that even in the absence of baryonic MACHOs there is still a 
limit, albeit much weaker, to $f_{\rm B}$ from LG dynamics.  The mass 
of baryons in the disk of the Galaxy provides a lower bound to 
$M_{\rm B}$ 
which is smaller by a factor of $\sim 3$ than the microlensing estimate 
we have used (\cite{fhp}).  Our lower bound to $f_{\rm B}$ would be 
reduced by this factor while our upper bound to $\Omega_0$ would be 
increased by the same factor.

In summary, if the observed microlensing events are the result of 
baryonic MACHOs in the Galaxy halo, then the dynamics of the LG may 
be used to infer a {\it lower} bound to the universal baryonic mass 
fraction: $f_{\rm B} > 0.29^{+0.18}_{-0.15}\, t_{10}^2$.  If primordial 
nucleosynthesis is used to provide an {\it upper} bound to the present 
baryonic density, we obtain an {\it upper} bound to the present total 
mass density: $\Omega_0  \alt 0.2$ (with an extereme upper bound
derived using nucleon-to-photon ratio based on the lithium abundance 
being $\Omega_0  \alt 0.47$).  

\acknowledgments

We thank S. Colombi, J. Felten, A. Gould, Kashlinsky, P. Sikivie, and 
A. Stebbins for useful discussions and helpful suggestions.  We reserve
special thanks for our referee, Jim Peebles, for his comments and 
questions which led to significant revision of and, we hope, improvement 
on our original manuscript.  
%We especially thank our referee, Jim Peebles, for his comments and 
%questions which led to significant revision of and, we hope, improvement 
%on our original manuscript.  
This work was supported at Ohio State 
by DOE grant DE-AC02-76ER01545.

\appendix
\section{Picture of infall after the removal of subclusters of galaxies.}

In the fitting of the infall in \S 4.1, all galaxies in the volume $r < 
3$ Mpc were considered.  However, galaxies which form gravitationally 
bound groups (or pairs) of their own, will have an additional peculiar 
velocity, unrelated to Local Group infall.  It is expected that the 
velocity dispersion will be smaller in modeling the infall onto the 
Local Group if only relatively isolated galaxies are considered.  
Therefore, it is interesting to repeat the analysis of \S 4.1 imposing 
such a selection.  The results of such an analysis are presented in 
this Appendix.

Galaxies are considered to form a group if the distance between them
is smaller than $0.5\, (M/M_{\rm M31})^{1/3}$ Mpc where M is the mass
of the galaxy pair and $M_{\rm M31}$ is the estimated mass of M31.  
Such groups were removed from the sample.  Since estimates for galaxy
masses were not presented in Karachentsev \& Makarov (1996), we took
them from  Peebles (1995, 1996).  We also removed a galaxy pair if the 
mass estimate was unavailable but the distance between the pair was smaller
than 0.2 Mpc.  The resulting infall diagram is shown in Fig.~\ref{rv2}, 
where the scale $r$ has been extended to 4~Mpc.  

{\plotone{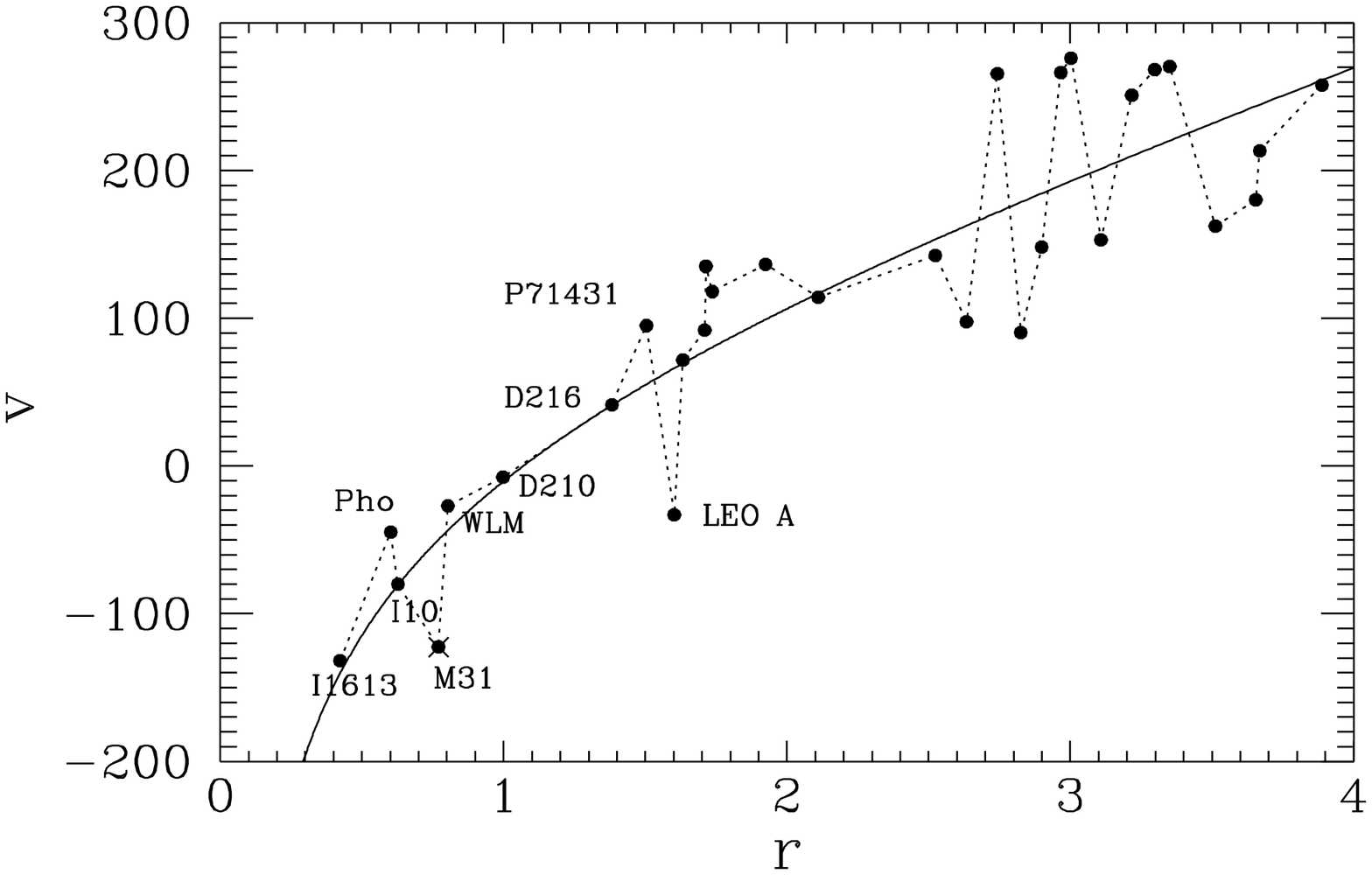}
\figcaption{
\baselineskip0.08cm
The same as Fig.~2, but after removal of close pairs of galaxies.  The 
solid line shows the best fit infall trajectory, $R=1.07$ Mpc, $h=0.71$. 
\label{rv2}}
}

The velocity dispersion did 
become significantly smaller; now we find $\sigma_v \approx 50$ km s$^{-1}$, 
while in \S 4.1 we obtained $\sigma_v \approx 70$ km s$^{-1}$.

In the analysis of \S 4.1 we had assumed that the Milky Way infalls to 
the common center of mass of galaxies in the volume $r < D$, while the 
velocity of the Galaxy with respect to those galaxies in the volume $D 
< r < 3$ Mpc was fitted.  This allowed us to account for the possible 
motion of the Local Group as a whole.  When we neglected the Local Group 
velocity, i.e., if $D = \infty$ was assumed, the quality of the fit became 
worse, i.e., the velocity dispersion increased.  It is interesting that 
the corresponding effect is very small in the present situation, when 
galaxy subclusters are removed.  Namely, by introducing the additional 
fitting parameters associated with the Local Group motion does reduce the
scatter of the data points around the infall trajectory, while the velocity 
dispersion (defined as $[\sum_i^N \Delta v_i^2/(N - N_{\rm par})]^{1/2}$, 
where $N$ is the number of data points and $N_{\rm par}$ is the number of 
fitting parameters) does not decrease.  Therefore, since the analysis is 
consistent with the assumption of negligible Local Group motion, we do not 
introduce the scale D here. 

Another measure of the robustness of this assumption can be obtained in 
the following way.  The velocity of the Milky Way can be found 
in fits with galaxy samples of varying depth.  
The direction and magnitude of the resulting best fit velocity can be 
compared (for each value of the depth) with the original assumption of 
infall to the common center of mass.  This trend is shown in Fig.~\ref{apex} 
for 
the angular coordinates of the velocity vector, and in Fig.~\ref{run_vel} 
for the 
amplitude of the velocity vector.  (For each subsequent point in 
Figs.~\ref{apex} 
and \ref{run_vel}, the number of galaxies in the fit is increased by one.)

\vspace{0.3cm}
{\plotone{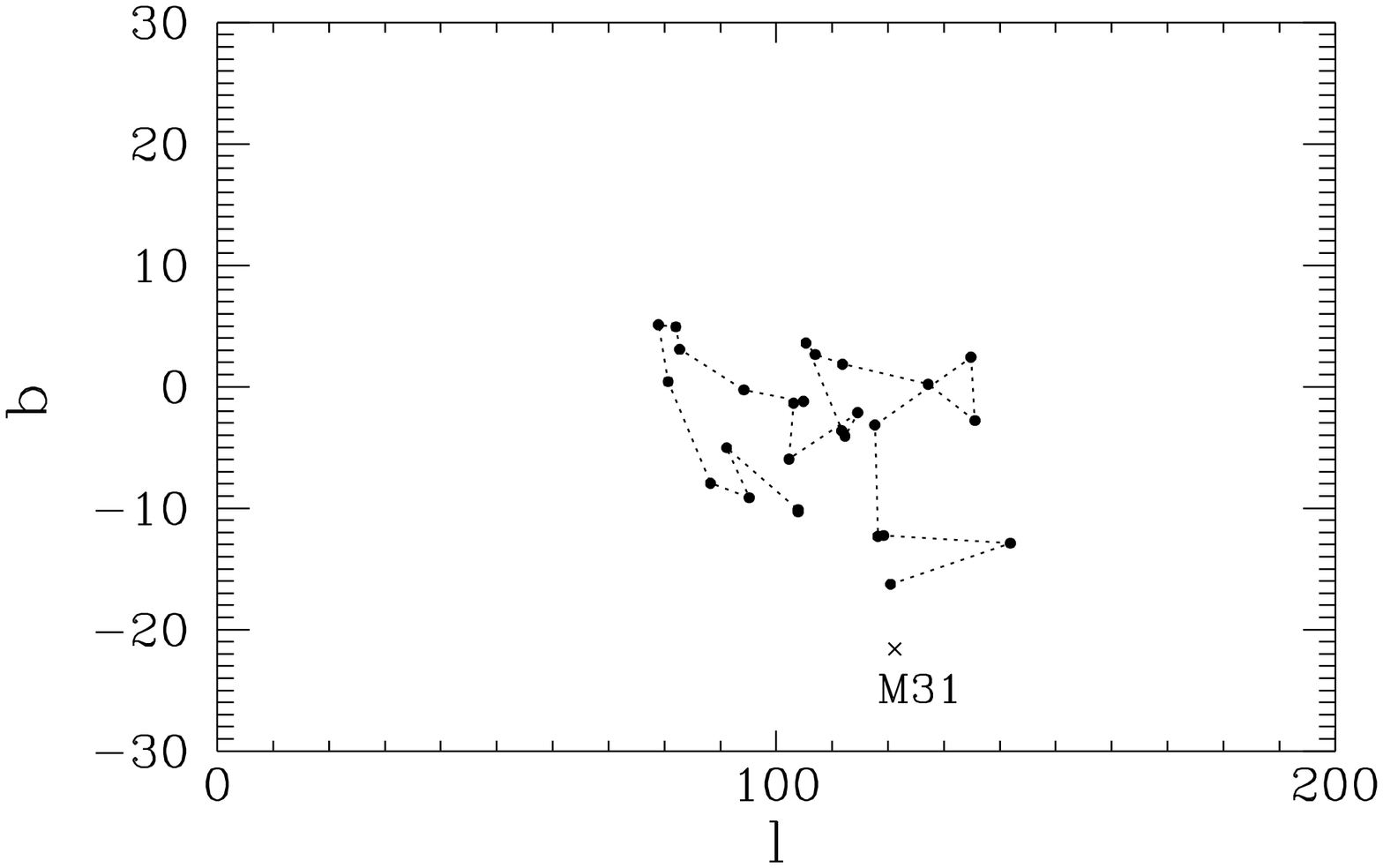}
\figcaption{
\baselineskip0.08cm
The Milky Way apex trend with respect to volumes of different depth.
The Andromeda (M31) position is shown by the diagonal cross.
\label{apex}}
}

\vspace{0.3cm}
{\plotone{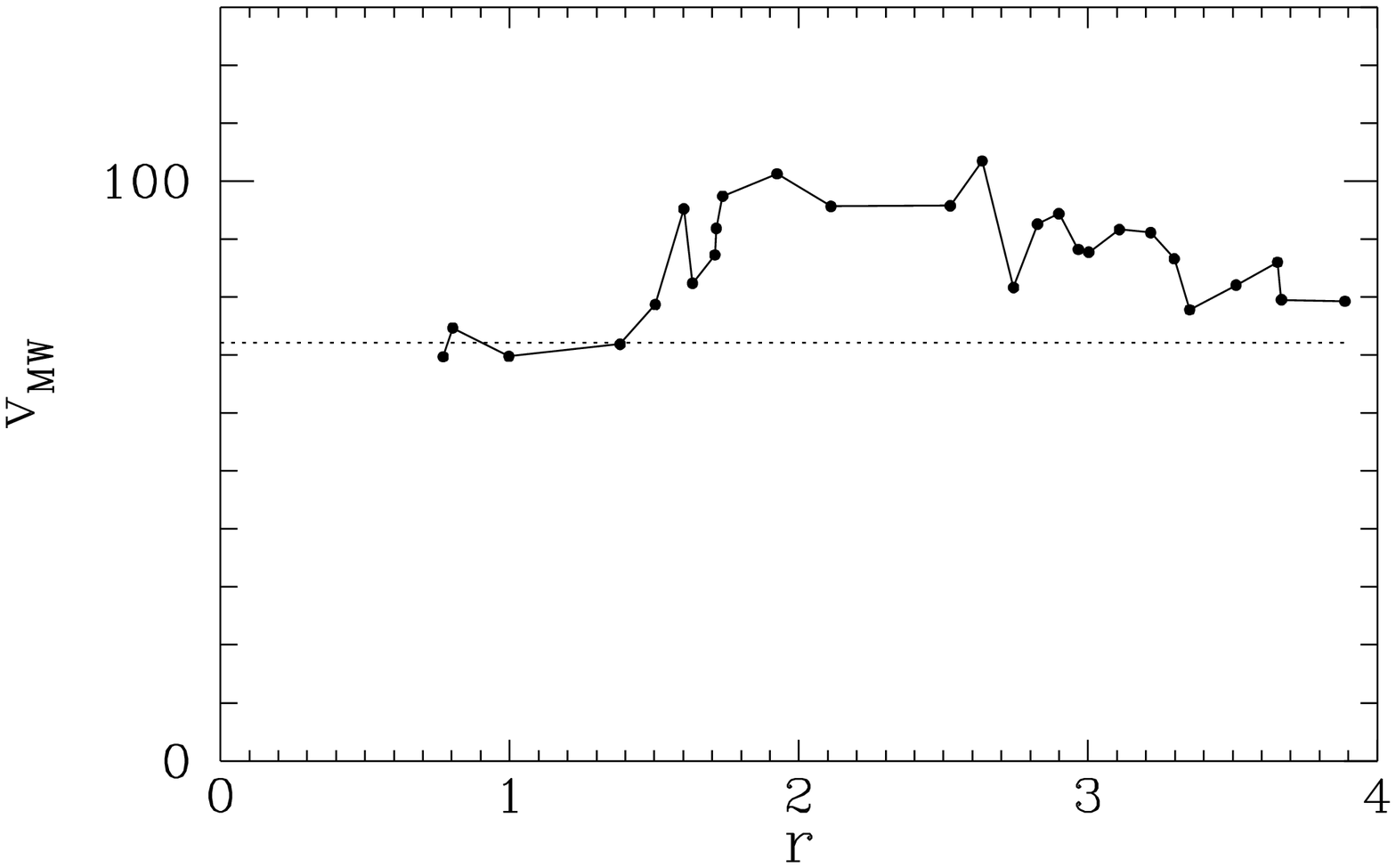}
\figcaption{
\baselineskip0.08cm
Magnitude of the Milky Way velocity towards an apex as a function of
the depth of the galaxy sample. The dotted line shows our predicted velocity.
%Expected velocity in our modeling is shown by the dotted line.
\label{run_vel}}
}
\vspace{0.3cm}

We see that with respect to nearby galaxies, the Milky Way does infall 
to M31 to good accuracy. Initially the direction is very close to the 
direction towards M31, and the magnitude is consistent with our assumption 
in \S 4.1 that at distances $r < 1.5 $ Mpc, $v_{\rm MW} = v_{\rm M31} 
(1+M_{\rm MW}/M_{\rm M31})^{-1}$.  When the number of galaxies in the 
sample increases (i.e., going to larger $r$), the direction of the Milky 
Way velocity drifts away from M31, and the magnitude of the velocity 
vector increases for $r > 1.5 $ Mpc.  However, at $r \approx 4$ Mpc, 
both the direction and magnitude turn back and, overall, in this volume 
the peculair velocity of the Local Group is not significant.

A similar analysis of the trend of the MW and LG apexes was done by 
Karachentsev \& Makarov (1996).  Their assumptions were different, 
however.  Instead of infall, Eqs. (5)-(\ref{chi2}), they assumed 
pure Hubble flow for galaxies at distances $r > 1.5$ Mpc, and $v=0$ 
for galaxies at distances $r < 1.5$ Mpc from the Milky Way (not from 
the center of mass of the Milky Way/M31 system).  Also, they included 
all close satellites in their fit.  For these reasons, our results, 
shown in Figs.~\ref{apex} and \ref{run_vel}, while consistent with 
theirs, are differing somewhat.

\end{document}